\documentclass[aps,prl,twocolumn]{revtex4}
\usepackage{amsmath,graphicx}

\begin{document}

\title{Optical coupling of fundamental whispering gallery modes in bi-spheres}
\author{L. I. Deych\footnote{E-mail:
  lev.deych@qc.cuny.edu}}
\affiliation{Physics Department, Queens College, City University of
New York, Flushing, New York 11367, USA.}
\author{C. Schmidt}
\author{A. Chipouline}
\author{T. Pertsch}
\affiliation{Institute of Applied Physics,
Friedrich-Schiller-University, Max-Wien-Platz 1, D-07743 Jena,
Germany.}
\author{A. T\"{u}nnermann}
\affiliation{Fraunhofer Institute for Applied Optics and Precision
Engineering, Albert-Einstein-Str. 7, D-07745, Jena, Germany.}

\begin{abstract}
What will happen if two identical microspheres, with fundamental
whispering gallery modes excited in each of them, become optically
coupled? Conventional wisdom based on coupled-mode arguments says
that two new modes, bonding and anti-bonding, with two split
frequencies would be formed. In this paper we demonstrate, using
exact multi-sphere Mie theory, that in reality an attempt to couple
two fundamental modes of microspheres  would result in a complex
multi-resonance optical response with the field distribution
significantly deviating from predictions of coupled-mode type
theories.
\end{abstract}

\maketitle

\paragraph{Introduction} Optical micro-resonators attract a great deal of
interest because of their unique properties such as small mode
volumes, high Q-factors, etc \cite{VahalaNature2002}. Individual
resonators of various shapes such as
microspheres\cite{BraginskyPhLetA1989},
microdisks\cite{AlmeidaNature2004},
microtoroids\cite{IlchenkoOL2001}, etc. have been carefully studied
and shown to have practically achievable Q-factors as large as
$10^{11} - 10^{13}$. Optical coupling between high-Q resonators due
to their evanescent fields draws attention because of unusual
optical effects that can arise in such systems and also from the
point of view of their potential for integrated optical circuits.
Simplest of such configurations is a double-resonator photonic
molecule suggested in\cite{BayerPRLOptMol1998} and studied both
experimentally and theoretically in many
papers\cite{RakovichPRA2004,{MukaiyamaPRL1999},{MiyazakiPRB2000}}.

In the case of microspheres, resonances are whispering gallery modes
(WGM) $|l,m,s\rangle$ characterized by angular, $l$, azimuthal, $m$,
and radial, $s$ quantum numbers. High values of $Q$ usually
corresponds to modes with $l\gg 1$, which, therefore, possess  a
high degree of degeneracy. The modes with the same orbital and
radial but different azimuthal numbers have the same
(complex-valued) frequency, but different spatial distributions. Of
greatest interest are so called "fundamental modes" with $|m|=l$,
$s=1$, whose field is tightly concentrated in the vicinity of the
equatorial plane and the surface of the sphere. It is assumed that
these modes can be selectively excited by coupling to a tapered
fiber\cite{ValhalaPRL2003}. When microspheres are arranged in
coupled structures the question of particular interest is the field
distribution resulting from coupling of fundamental modes of
individual spheres. In the framework of coupled-mode
theories\cite{HausJLWT1987,{LittleJLWT1997}} or perturbation
approaches\cite{PovinelliOpticsExpress2005Forces} this distribution
is often described as a linear combination of phase-matched modes
with $m=l$ in one sphere and $m=-l$ in the other, with the coupling
strength characterized by an overlap integral of the respective
modal
functions\cite{ShopovaPRA2005IndTransp,{PovinelliOpticsExpress2005Forces}}.

\begin{figure}
  \includegraphics[width=3.0in]{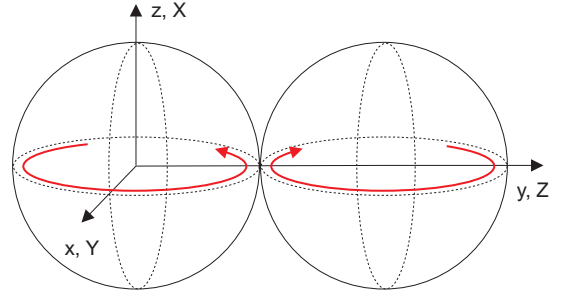}\\
    \caption{Configuration of a two-sphere system and possible coordinate systems. Arrowed lines show schematically two single-sphere fundamental modes}
    \label{fig:layout}
\end{figure}

However, it will be shown below that such field configurations are
not consistent with the symmetry of the system if the deviations of
the spheres from the ideal shape is negligible.  More accurately, we
demonstrate, analytically and numerically, that if one excites a
fundamental mode $|L,L,1\rangle$ in one or both uncoupled spheres
and bring them together for optical coupling, the ensuing violation
of the complete spherical symmetry of the system results in a field
configuration containing linear combination of all initially
degenerate modes with $|m|\le L$. The optical coupling removes the
degeneracy of these modes so that each one of them gives rise to a
couple of resonance peaks at its own frequencies producing a
complicated multi-peak optical response.  Moreover, because of the
coupling between modes with different $l$ the field of the bi-sphere
contains a significant admixture of modes $|l,m,s\rangle$ with $l\ne
L$ and $s>1$.
\paragraph{Fundamental modes and coordinate systems.} The system considered
in the paper consists of two identical dielectric spheres of radius
$R$ and refractive index $n$ positioned at a distance $d \geq 2R$
between their centers (see Fig.\ref{fig:layout}). We assume that the
field of the system is monochromatic with frequency $\omega$ and
separate it into a combination of incident, scattered, and internal
fields. Using a standard multi-sphere Mie
theory\cite{FullerApplOpt1991,Mishchenko_book2002,MiyazakiPRB2000}
we present all these fields as linear combinations of single-sphere
vector spherical harmonics (VSH).
In this representation, the fields are characterized by expansion
coefficients $\zeta_{l,m}^{(i)}$, $ a_{l,m}^{(i)}$, $c_{l,m}^{(i)}$
for incident, scattered, and internal fields of TM polarization,
respectively, and $\eta_{l,m}^{(i)}$, $b_{l,m}^{(i)}$,
$d_{l,m}^{(i)}$ for TE polarization. Here index $i$ enumerates
spheres and  takes only two values $i=1,2$. Using Maxwell boundary
conditions and the addition theorem for the
VSH\cite{CruzanApplMath1962,{SteinApplMath1961}} one can derive a
system of equations relating scattering coefficients $a_{l,m}^i$ and
$b_{l,m}^i$ to the coefficients of the incident field
$\zeta_{l,m}^{i}$ and~$\eta_{l,m}^{i}$:
\begin{equation}
\begin{split}
a_{l,m}^{i}=\alpha(x)_{l}^{(N)}\left\{\zeta_{l,m}+\sum\limits_{j\neq
i}\sum\limits_{l^\prime,m^\prime}\left[a_{l^\prime,m^\prime}^{j}A_{l,m}
^{l^\prime,m^\prime}\left(x,\mathbf{r}_j-\mathbf{r}_i\right)\right.\right.\\
 +\left.\left.b_{l^\prime,m^\prime}^{j}B_{l,m}^{l^\prime,m^\prime}\left(x,\mathbf{r}_j-\mathbf{r}_i\right)\right]\right\}\label{eq:a_coeff_expan}
 \end{split}
\end{equation}
\begin{equation}
\begin{split}
b_{l,m}^{i}=\alpha(x)_{l}^{(M)}\left\{\eta_{l,m}+\sum\limits_{j\neq
i}\sum\limits_{l^\prime,m^\prime}\left[b_{l^\prime,m^\prime}^{j}A_{l,m}
^{l^\prime,m^\prime}\left(x,\mathbf{r}_j-\mathbf{r}_i\right)\right.\right.\\
+\left.\left.
a_{l^\prime,m^\prime}^{l}B_{l,m}^{l^\prime,m^\prime}\left(x,\mathbf{r}_j-\mathbf{r}_i\right)\right]\right\}\label{eq:b_coeff_expan}
\end{split}
\end{equation}
where  $\mathbf{r}_i$ is the position vector of $i$-th sphere,
$\alpha_l^{(N)}$ and $\alpha_l^{(M)}$ are single sphere Mie
scattering coefficients for $TM$ and $TE$ polarizations
respectively.  These coefficients have poles at the specific values
of the dimensionless frequency parameter $x$, defined as
$x=nR\omega/c$, which determine frequency and the spectral width of
the single-sphere WGM resonances. Explicit expressions for the
scattering coefficients as well as for translational coefficients
$A_{l,m}^{l^\prime,m^\prime}\left(\mathbf{r}_j-\mathbf{r}_i\right)$
and
$B_{l,m}^{l^\prime,m^\prime}\left(\mathbf{r}_j-\mathbf{r}_i\right)$,
which describe optical coupling between the spheres can be found,
for instance in
Ref.\cite{Mishchenko_book2002,FullerApplOpt1991,MiyazakiPRB2000}.

Our goal is to study effects of  optical coupling on fundamental
modes of single spheres. Classification of WGMs as fundamental,
however, is linked to a particular coordinate system: modes with
$|m|=l$ are associated with the plane perpendicular to polar axis
$z$.
In order to achieve strong coupling between fundamental modes in a
bi-sphere one needs to choose a polar axes such that the respective
equatorial plane would contain centers of both spheres. This
coordinate system is labeled by lower case letters in
Fig.\ref{fig:layout}. 
However, this coordinate system is not consistent with the symmetry
of the bi-sphere, so that its modes defined using associated
spherical coordinates cannot be characterized by azimuthal number
$m$.  Formally, it follows from the presence of non-diagonal in $m$
elements in the translation coefficients. Consequently, the
configuration of the fields in this system cannot be presented as a
combination of the modes with $|m|=l$ contrary to the assumption of
the coupled-mode theories.
In a coordinate system with polar axis along the line of symmetry of
the system ($XYZ$ system in Fig.\ref{fig:layout}) the translation
coefficients are diagonal in $m$ and the normal modes of the coupled
system can again be classified according to the azimuthal
number\cite{MiyazakiPRB2000,{DeychRoslyakPRE2006}}. However, the
field distribution corresponding to the fundamental mode of the
$xyz$ coordinate system cannot be characterized by a VSH with a
single $m$ in the $XYZ$ system. In order to describe the same field
distribution in the new coordinates one has to use transformation
properties of VSH\cite{Mishchenko_book2002} and express a single
$|L,L,1\rangle$ mode of $xyz$ system as a linear combination of all
$|L,m,1\rangle$ modes of the $XYZ$ system. In order to reproduce
such field configuration in a single sphere we choose the incident
field to be TE polarized with expansion coefficients of the form
\begin{equation}\label{eq:incident_new}
\eta_{l,m}^{(i)}=\delta_{lL}\delta_{i1}R_{Lm},\hskip 5pt
R_{Lm}=\frac{(-i)^L}{2^L}\sqrt{\displaystyle{\frac{(2L)!}{(L+m)!(L-m)!}}}
\end{equation}
which is dictated by the transformation properties of VSH.
\paragraph{Fundamental modes in the bi-sphere}
Since translational coefficients in the new coordinate system become
diagonal in $m$, expansion coefficients in
Eq.(\ref{eq:a_coeff_expan}) and (\ref{eq:b_coeff_expan}) with
different values of the azimuthal number become independent and can
be solved separately. In what follows we will also neglect the
cross-polarization terms, described by coefficients
$B_{lm}^{l^\prime m}$, which are usually much smaller than
$A_{lm}^{l^\prime m}$\cite{MiyazakiPRB2000}. Then we need only to be
concerned with coefficients $b_{lm}$ and
Eq.(\ref{eq:b_coeff_expan}). In order to understand qualitatively
the effect of the coupling on the field distribution in the
bi-sphere we will first solve this equation in so called single-mode
approximation neglecting interaction between modes with different
$l$-numbers (see details in
Ref.\cite{MiyazakiPRB2000,{DeychRoslyakPRE2006}}). Expression for
the scattered field in this approximation can be presented in the
following form
\begin{equation}\label{eq:single_mode}
\begin{split}
\mathbf{E_{s}}=\frac{1}{2}\sum_mR_{Lm}\left[\frac{\mathbf{M}_{Lm}(\mathbf{r}-\mathbf{r}_1)+\mathbf{M}_{Lm}(\mathbf{r}-\mathbf{r}_2)}{\alpha_L^{-1}-A_{Lm}^{Lm}(\mathbf{r}_2-\mathbf{r}_1)}\right.\\
\left.\frac{\mathbf{M}_{Lm}(\mathbf{r}-\mathbf{r}_1)-\mathbf{M}_{Lm}(\mathbf{r}-\mathbf{r}_2)}{\alpha_L^{-1}+A_{Lm}^{Lm}(\mathbf{r}_2-\mathbf{r}_1)}\right],
\end{split}
\end{equation}
which gives a clear physical picture of the phenomenon under
consideration. This expression describes an optical response with
resonances at two sets of frequencies: one is given by the zeroes of
$\alpha_L^{-1}-A_{Lm}^{Lm}$ and the other by the zeroes of
$\alpha_L^{-1}+A_{Lm}^{Lm}$. The role of coupling between spheres is
described by translation coefficients $A_{Lm}^{Lm}$, which shifts
resonant frequency from the single sphere values by {\it different
amounts for different values of} $m$. As a result terms with
different azimuthal numbers resonate at different frequencies so
that one ideally could expect $2(L+1)$ resonance peaks in the
optical response of the system contrary to the coupled-mode theory
expectation of just two resonances. The number of observable peaks
is smaller because frequency shifts for higher values of $|m|$ are
small and resonances from respective terms may not be
distinguishable.

One can notice that the numerators of terms in
Eq.(\ref{eq:single_mode}) corresponding to different values of $m$
have a form of symmetric and antisymmetric combinations of
respective single sphere modes, which is typical for the coupled
mode approaches. In our consideration, however, this combination
appears  for every $m$-component constituting the fundamental mode
separately, rather than for the entire single sphere mode. 

Additional deviations from the coupled-mode approximation appear due
to coupling between modes with different values of the angular
number $l$, which is neglected in the single-mode approximation.
This approximation can be improved by perturbation methods described
in Ref.\cite{MiyazakiPRB2000,{DeychRoslyakPRE2006}}, in this paper,
however, we will rely on exact numerical treatment of
Eq.(\ref{eq:b_coeff_expan}).
\begin{figure}
  \includegraphics[width=4in]{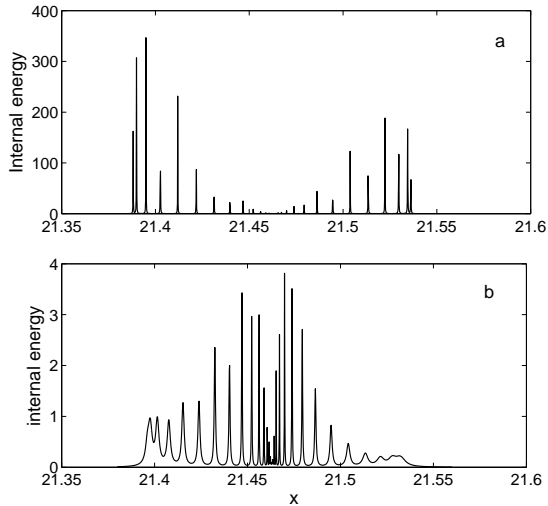}\\
  \caption{Frequency dependence of the internal energy of the bi-sphere in the vicinity of single sphere resonance frequency x=21.46 corresponding to a TE mode $|29,m,1\rangle$.  (a) single-mode approximation; (b) numerical calculations including all modes with $l\leq 29$}\label{fig:energy}
\end{figure}
For numerical analysis of Eq.(\ref{eq:b_coeff_expan}) with incident
field in the form of Eq.(\ref{eq:incident_new}) we chose $L=29$ and
spectral interval in the vicinity of resonance frequency of the mode
with radial number $s=1$, $x_{29,1}= 21.46$. We solve this equation
using matrix inversion and including all coefficients with $l\leq
L$. The limitations for the number of modes from above is made only
to shorten the required time of calculation, but it does not affect
the results significantly: we extended the number of included modes
up to $l=48$ for several values of the dimensionless frequency $x$
and observed good convergence of the results with only small
quantitative changes resulting from contribution of the modes with
$l>L$. Taking into account the coefficients with $l<L$, however, is
significant because some modes with smaller values of $l$ might have
spectral overlap with the main mode $l=L$ and their contribution can
be resonantly enhanced\cite{MiyazakiPRB2000,{DeychRoslyakPRE2006}}.
Using found scattering coefficients we calculate expansion
coefficients of the internal field $d_{l,m}$:
\begin{equation}
d_{lm}^{(i)}=\frac{i}{x}\frac{1}{j_l(x)[nxj_l(nx)]^\prime -
j_l(nx)[xj_l(x)]^\prime}b_{lm}^{(i)}
\end{equation}
where $j_l(x)$ is the spherical bessel function and
$[zj_l(z)]^{\prime}$ means differentiation with respect to $z$.
Knowing coefficients $d_{lm}$ we can find total energy of the field
concentrated inside spheres as a function of frequency, which is
best suited to characterize optical response of our system in the
spectral range of high-Q WGMs\cite{MiyazakiPRB2000}.
Fig.\ref{fig:energy} presents the spectra of the internal energy
obtained by two procedures: (a) calculations based on the
single-mode approximation, Eq.(\ref{eq:single_mode}), and (b)
multi-mode numerical calculations described above. One can see that
while the single-mode model reproduces the multi-resonance optical
response, it deviates strongly from exact numerical calculations in
positions and heights of the respective peaks. The most striking
difference is significant lowering of the heights of the peaks when
inter-mode coupling is taken into account, which indicates a large
(three orders of magnitude) reduction of Q-factors of some of the
resonances due to coupling to low-Q modes with $l\leq L$.

\begin{figure}
  \includegraphics[width=3.5in]{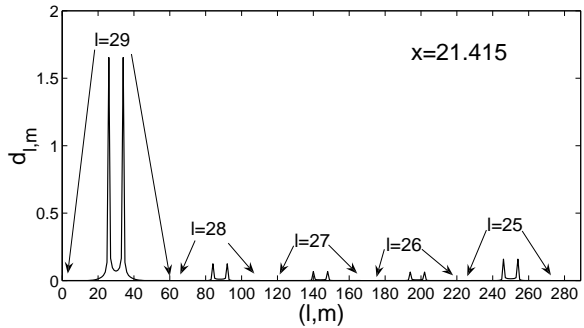}\\
    \caption{Internal field coefficients for all values of $l$ and $m$ for $l\geq 25$. Arrows indicate the regions corresponding to coefficients with given $l$ and varying $m$.}\label{fig:d_coeff}
\end{figure}

In order to further elucidate the role of the inter-mode coupling we
consider the coefficients of the internal field $d_{lm}$ for all
values of $0<l\leq 29$ and respective values of $|m|\leq l$. In
order to visualize the results we present pairs $l,m$ as an
one-dimensional array ordered according to the following rule
$(29,-29), (29,-28), \ldots, (29,0), (29,1)$, $\ldots, (29,29),
(28,-28), \ldots$ and plot coefficients $d_{lm}$ versus a number of
the respective pair in this array. Fig.~\ref{fig:d_coeff} presents
the results of these calculations for $l\geq 25$ for one of the
resonance frequencies, $x=21.415$. One can see that the largest
values of the coefficients correspond to the main angular number
$l=29$, which as a function of $m$ shows two large symmetric peaks
for $m=\pm 4$. The positions of the peaks are determined by the
dependence of the resonant frequency on the azimuthal number: the
resonance at the chosen frequency results from the component of the
internal field with $|m|=4$. In addition to main coefficients
corresponding to $l=29$ the plot reveals presence of other
coefficients as well. The second largest coefficient corresponds to
$l=25$, which is in agreement with the fact that a frequency of the
mode $|25,m,2\rangle$ is almost in resonance with our main mode,
$|29,m,1\rangle$\cite{MiyazakiPRB2000}. While individual
contributions from modes with $l<29$ appears to be small, their
cumulative effect is significant as evidenced by
Fig.\ref{fig:energy}. The role of the inter-mode coupling can be
even more important if one attempts to excite other fundamental
modes such as TE mode $|39,39,1\rangle$. This mode interacts
resonantly with TE modes,
$|34,m,2\rangle$\cite{MiyazakiPRB2000,{DeychRoslyakPRE2006}} because
of their large Q-factors  and significant spectral overlap with
$|39,39,1\rangle$. Deviations from single mode approximation in this
case will be even more drastic.
\paragraph{Conclusion} In this paper we studied effects of optical coupling on the fundamental modes excited
in two spherical microresonators. We found that the optical response
of this system under these excitation conditions is characterized by
a rich spectrum with multiple resonances in contrast with just two
peaks predicted by simplified coupled-mode type approaches. These
multiple resonances arise due to the fact that strongly coupled
fundamental modes of single spheres do not constitute a normal mode
of a bi-sphere and must, therefore, be described by a linear
combination of vector spherical harmonic with different azimuthal
numbers.  Optical coupling removes degeneracy between resonances
with different $m$, causing, therefore, different $m$-components of
the field to resonate at different frequencies and producing
multi-peak spectrum. We also showed that coupling between modes with
different angular numbers $l$ also significantly affects spectrum
and the field distribution in the bi-sphere.

The question arises, however, how do our results agree with
observations of bonding and anti-bonding orbitals with two split
frequencies reported in many works? First of all, modes observed in
many experiments were true \textbf{normal modes} of the bi-spheres,
characterized by a well defined azimuthal number $m$
\cite{RakovichPRA2004}). These modes, however, are not coupled
\textbf{fundamental} modes and results of this work do not apply to
those experiments.

In order to observe effects presented in this paper one needs to
excite a fundamental mode in a single sphere concentrated in a plane
that would contain centers of both spheres, and measure optical
response of the bi-sphere under the same excitation conditions. This
might be possible with the use of tapered fiber excitation
techniques\cite{ValhalaPRL2003}, one, however, should be aware of
 sensitivity of the multiple peak response to the strength of
coupling and Q-factors of the modes. Increasing distance between the
spheres in our numerical simulations, we observed how multiple peaks
collapse to form a double peak structure. It would be a mistake,
however, to associate these two peaks with the bonding and
anti-bonding states of the coupled-mode theory because these peaks
arise due to smearing of original multiple peaks rather than due to
splitting of a single sphere resonance. The width of those peaks
would represent an inhomogeneous broadening caused by inability to
resolve the fine structure the spectrum rather than the radiative
decay rate of respective modes.

One can see, therefore, that the picture of two coupled fundamental
modes, which is used widely in modeling of optical networks can be
misleading, and we hope that this work will serve as a warning that
one needs to exercise caution when analyzing experimental data and
developing theoretical models based on this picture.

\end{document}